\documentclass[fleqn,twoside]{article}
\usepackage{espcrc2,epsf,latexsym}

\newcommand{\half}{\mbox{\small $\frac{1}{2}$}}          
\newcommand{\third}{\mbox{\small $\frac{1}{3}$}}         
\newcommand{\twothird}{\mbox{\small $\frac{2}{3}$}}      
\newcommand{\msbar}{\mbox{\tiny $\overline{MS}$}}        
\newcommand{\NS}{\mbox{\tiny $N\!S$}}                    

\def\lsim{\mathrel{\rlap{\lower4pt\hbox{\hskip1pt$\sim$}}
    \raise1pt\hbox{$<$}}}                
\def\gsim{\mathrel{\rlap{\lower4pt\hbox{\hskip1pt$\sim$}}
    \raise1pt\hbox{$>$}}}                

\title{
       \vspace{-3.65cm}                                     %
       {\normalsize DESY 03-183}      \\[-0.2cm]            
       {\normalsize Edinburgh 2003/19}\\[-0.2cm]            
       {\normalsize LTH 610}          \\[-0.2cm]            
       {\normalsize LU-ITP 2003/26}   \\[-0.2cm]            
       {\normalsize November 2003}    \\                    
       \vspace{1.32cm}                                      
       Structure functions and form factors close to the chiral 
       limit from lattice QCD%
            \thanks{Talk given by R. Horsley at LHP2003,
                    Cairns, Australia.}}                    

\author{T. Bakeyev%
           \address{Joint Institute for Nuclear Research,
                    RU-141980 Dubna, Russia},
        D. Galletly%
           \address{School of Physics,
                    University of Edinburgh, Edinburgh EH9 3JZ, UK},
        M. G\"ockeler%
           \address{Institut f\"ur Theoretische Physik, Universit\"at
                    Leipzig, D-04109 Leipzig, Germany}${}^,\hspace*{-0.05in}$
           \address{Institut f\"ur Theoretische Physik,
                    Universit\"at Regensburg, D-93040 Regensburg, Germany},
        M. G\"urtler%
           \address{John von Neumann Institute NIC/DESY Zeuthen,
                    D-15738 Zeuthen, Germany},
        R. Horsley%
           $^{\rm b}$,
        B. Jo\'o%
           $^{\rm b}$,
        A.~D. Kennedy%
           $^{\rm b}$,
        B. Pendleton%
           $^{\rm b}$,
        H. Perlt%
           $^{\rm d,c}$,
        D. Pleiter%
           $^{\rm e}$,
        P.~E.~L. Rakow%
           \address{Department of Mathematical Sciences, 
                    University of Liverpool, Liverpool L69 3BX, UK},
        G. Schierholz%
           $^{\rm e,}$%
           \address{Deutsches Elektronen-Synchrotron DESY,
                    D-22603 Hamburg, Germany},
        A. Schiller%
           $^{\rm c}$,
        T. Streuer%
           $^{\rm e,}$%
           \address{Institut f\"ur Theoretische Physik,
                    Freie Universit\"at Berlin, D-14196 Berlin, Germany}
        and
        H. St\"uben%
            \address{Konrad-Zuse-Zentrum f\"ur Informationstechnik Berlin,
                    D-14195 Berlin, Germany}
        -- {\it QCDSF--UKQCD} Collaboration }

\begin{document}

\begin{abstract}
Results for nucleon matrix elements (arising from moments
of structure functions) and form factors from a mixture of runs
using Wilson, clover and overlap fermions (both quenched and unquenched)
are presented and compared in an effort to explore the size of 
the chiral `regime', lattice spacing errors and quenching artefacts.
While no run covers this whole range of effects the partial results
indicate a picture of small lattice spacing errors, small quenching
effects and only reaching the chiral regime at rather light quark masses.
\end{abstract}

\maketitle

\setcounter{footnote}{0}


\section{INTRODUCTION}

Chiral extrapolations of lattice data to the physical pion mass,
the continuum or $a \to 0$ limit and removal of the `quenched approximation'
remain major sources of systematic uncertainty in the determination of
hadron matrix elements (see eg \cite{gockeler02a}). A study of these
quantities is necessary to understand how QCD binds quarks and gluons to form
hadronic states and to give an explanation about how particle
mass and spin arises. Due to the importance of proton/neutron scattering 
and DIS experiments most is known about the nucleon.
In this talk we shall detail our recent progress in studying
these problems from the lattice or numerical perspective
in particular with regard to the chiral extrapolation.
We have computed $v_n \equiv \langle x^{n-1} \rangle$,
for $n = 2$, $3$ and $4$ related to the three lowest unpolarised
moments of the $F_2$ nucleon structure function, which are given by
\begin{eqnarray}
 \lefteqn{
    \langle N(\vec{p}) | \left[ \widehat{\cal O}^{(q)\{ \mu_1\cdots\mu_n \} }
                     - \mbox{\rm Tr} 
                      \right] | N(\vec{p}) \rangle^{\msbar} = }
             & &                                      \nonumber  \\
             & &  \hspace*{0.75in}
                    2 v^{(q){\msbar}}_n  [p^{\mu_1} \cdots p^{\mu_n}
                     - \mbox{\rm Tr}] \,,
\end{eqnarray}
where
\begin{eqnarray}
 {\cal O}^{(q)\mu_1\cdots\mu_n}
   &\equiv& \mbox{i}^{n-1} \overline{q}\gamma^{\mu_1}
            \stackrel{\leftrightarrow}{D^{\mu_2}} \cdots
            \stackrel{\leftrightarrow}{D^{\mu_n}}q \,,
\end{eqnarray}
($q = u, d$). In this report we shall restrict attention to the
$n=2$ moment only. We shall also consider here form factors arising
from lepton-nucleon scattering,
\begin{eqnarray}
  \lefteqn{\langle  N(\vec{p}^{\,\prime}) |
           \widehat{\cal V}^{(\twothird u - \third d)}_\mu(\vec{q})
                          | N(\vec{p}) \rangle = }
         &                                    \nonumber \\
         &  \hspace*{0.075in}
              \overline{u}_N(\vec{p}^{\,\prime},\vec{s}^{\,\prime})
              \left[ \gamma_\mu 
                     F^p_1 +
                     \mbox{i}\sigma_{\mu\nu}
                     {q^\nu \over 2m_N} F^p_2
              \right] u_N(\vec{p},\vec{s}) \,,
\label{formfactor}
\end{eqnarray}
where ${\cal V}^{(q)}_\mu \equiv \overline{q}\gamma_\mu q$
is the vector current and the momentum transfer $q = p^{\,\prime} - p$.

Specifically we shall consider a variety of actions and quark mass
ranges,
\begin{enumerate}
   \item Wilson fermions at one fixed lattice spacing,
         $a^{-1} \sim 2.12\mbox{GeV}$ in the quenched approximation
         at pseudoscalar masses down to $\sim 310\mbox{MeV}$,
         in order to try to match to chiral perturbation theory ($\chi$PT).
   \item $O(a)$-improved Wilson fermions (`clover fermions') 
         in the quenched approximation at three lattice spacings
         $a^{-1} \sim 2.12$ -- $3.85 \mbox{GeV}$ with pseudoscalar
         masses between $580\mbox{MeV}$ and $1200\mbox{MeV}$
         to check finite lattice artefacts.
   \item Unquenched clover fermions at pseudoscalar masses down to
         $\sim 560\mbox{MeV}$ in order to see if there are any
         discernable quenching effects.
   \item Overlap fermions, in the quenched
         approximation at one lattice size $a^{-1} \sim 2.09\mbox{GeV}$
         down to pseudoscalar masses of about $440\mbox{MeV}$.
         These have a chiral symmetry even with finite lattice spacing
         and hence have better chiral properties than either Wilson
         or clover fermions.
\end{enumerate}
Note that the physical pion mass is about $m_\pi \sim 140\mbox{MeV}$
and we use $r_0 = 0.5\mbox{fm} \equiv (394.6\mbox{MeV})^{-1}$
to set the scale.

What must one achieve to be able to compare numerical results with QCD?
One hopes that after taking the continuum limit, we can match
our results to known chiral perturbation theory and then take the
limit $m_{ps} \to m_\pi$. Practically we might thus expect that for
a quantity $Q$ of interest
\begin{eqnarray}
   Q = F_\chi^Q(r_0m_{ps}) + O((a/r_0)^n) \,.
\end{eqnarray}
$F_\chi^Q(r_0m_{ps})$ describes the (chiral) physics and the
discretisation errors are $O(a^n)$ where $n=1$ for Wilson fermions
and $n=2$ for clover and overlap fermions. The $O(a)$ errors may
be split into terms that remain in the chiral limit, $O(a/\Lambda)$,
and $O(am_q) \sim O(ar_0m_{ps}^2)$. While $O(a)$ improvement
will remove both these terms, near the pion mass this second
set of lattice errors will be negligible anyway, so we do not have
to worry about this term. Thus the variant we shall try here for clover
fermions is
\begin{eqnarray}
   Q = F_\chi^Q(r_0m_{ps}) + d^Q_a (a/r_0)^2 + d^Q_m ar_0 m_{ps}^2 \,,
\label{practical}
\end{eqnarray}
where we shall take $d^Q_a$ and $d^Q_m$ to be constant.

Naively one might expect a Taylor series expansion for $F^Q_\chi$
to be sufficient, ie
\begin{eqnarray}
   F^Q_\chi(x) = F^Q_\chi(0) + c^Qx^2 + \ldots \,.
\label{taylor}
\end{eqnarray}
Over the last few years expressions for $F^Q_\chi$ have been found
\begin{eqnarray}
   F_\chi^{m_N}(x)   &\hspace*{-0.075in}=\hspace*{-0.075in}&
                      r_0m_N(0) + [c^{m_N}_{\half}x] + c_1^{m_N}x^2 + \ldots
\label{chiPT}
                                                                \\
   F_\chi^{v_n}(x)   &\hspace*{-0.075in}=\hspace*{-0.075in}&
                       v_n(0)\left( 1 + 
                       c_1^{v_n}x^2\ln (x/r_0\Lambda_\chi)^2 \right) + \ldots
                                                     \nonumber
\end{eqnarray}
In general leading order $\chi$PT introduces non-analytic terms
to the naive Taylor expansion of $F^Q_\chi$, either odd powers,
such as $x$ or $x^3$ in $F_{\chi}^{m_N}$ or logarithmic as in $F_\chi^{v_n}$.
Additional terms may appear when using the quenched approximation;
these are shown in the above equations inside square brackets.
The chiral scale, $\Lambda_\chi$, is usually taken to be $\sim 1\mbox{GeV}$.
Expressions are known for some of the coefficients in the above expansions
(using for example the results of
\cite{labrenz93a,bernard93a,detmold01a,arndt01a,chen01a})
and rough numerical estimates give
\begin{eqnarray}
   c^{m_N}_{\half} &\hspace*{-0.050in}=\hspace*{-0.050in}&
                  -{3\pi \over 2}(D-3F)^2\delta \sim -0.5 \,,
                                                     \nonumber \\
   c^{m_N}_1       &\hspace*{-0.050in}=\hspace*{-0.050in}&
                  2(b_D - 3b_F) \sim  1.3                 \,,
\label{chiPT_numerical}
\end{eqnarray}
and similarly
\begin{eqnarray}
   c^{v_n}_1  = -{3g_A^2+1 \over (4\pi r_0f_\pi)^2} \sim - 0.66 \,.
\end{eqnarray}
For quenched QCD an expression for $c^{v_n}_1$ in terms of $D$ and $F$
can be found in \cite{chen01b} giving an approximate result of
\begin{eqnarray}
   c^{v_n}_1  \sim - 0.28 \,.
\label{numerical_c2}
\end{eqnarray}


\section{LATTICE DETAILS}


\subsection{Run parameters}

Details of the statistics and parameter values used in the
simulations for the nucleon matrix elements are given in 
tables~\ref{table_qW}, \ref{table_qC}, \ref{table_uqC} and \ref{table_qOLW},
\begin{table}[ht]
   \begin{center}
      \begin{tabular}{l|l|l|l}
          $\beta$    & $\kappa$ & Volume        & Confs.  \\
         \hline
               6.0   & 0.1515 & $16^3\times 32$ & $O(980)$  \\
               6.0   & 0.1530 & $16^3\times 32$ & $O(1130)$ \\
               6.0   & 0.1550 & $16^3\times 32$ & $O(1360)$ \\
         \hline
               6.0   & 0.1550 & $24^3\times 32$ & $O(220)$ \\
               6.0   & 0.1558 & $24^3\times 32$ & $O(220)$ \\
               6.0   & 0.1563 & $24^3\times 32$ & $O(220)$ \\
         \hline
               6.0   & 0.1563 & $32^3\times 48$ & $O(250)$ \\
               6.0   & 0.1566 & $32^3\times 48$ & $O(530)$ \\
      \end{tabular}
   \end{center}
   \caption{\footnotesize{Parameters for quenched Wilson fermions 
            with Wilson glue. (The smallest pseudoscalar mass is
            $\sim 310\mbox{MeV}$.)}}
   \vspace*{-0.30in}
   \label{table_qW}
\end{table}
\begin{table}[ht]
   \begin{center}
      \begin{tabular}{l|l|l|l}
          $\beta$    & $\kappa$ & Volume        & Confs.  \\
         \hline
               6.0   & 0.1320 & $16^3\times 32$ & $O(445)$ \\
               6.0   & 0.1324 & $16^3\times 32$ & $O(560)$ \\
               6.0   & 0.1333 & $16^3\times 32$ & $O(560)$ \\
               6.0   & 0.1338 & $16^3\times 32$ & $O(520)$ \\
               6.0   & 0.1342 & $16^3\times 32$ & $O(735)$ \\
         \hline
               6.2   & 0.1333 & $24^3\times 48$ & $O(300)$ \\
               6.2   & 0.1339 & $24^3\times 48$ & $O(300)$ \\
               6.2   & 0.1344 & $24^3\times 48$ & $O(300)$ \\
               6.2   & 0.1349 & $24^3\times 48$ & $O(470)$ \\
         \hline
               6.4   & 0.1338 & $32^3\times 48$ & $O(220)$ \\
               6.4   & 0.1342 & $32^3\times 48$ & $O(120)$ \\
               6.4   & 0.1346 & $32^3\times 48$ & $O(220)$ \\
               6.4   & 0.1350 & $32^3\times 48$ & $O(320)$ \\
               6.4   & 0.1353 & $32^3\times 64$ & $O(260)$ \\
      \end{tabular}
   \end{center}
   \caption{\footnotesize{Parameters for quenched clover fermions 
            with Wilson glue.}}
   \vspace*{-0.30in}
   \label{table_qC}
\end{table}
\begin{table}[ht]
   \begin{center}
      \begin{tabular}{l|l|l|l|l}
          $\beta$    & $\kappa_{sea}$& Volume   & Trajs.  &  Group \\
         \hline
               5.2   & 0.1342 & $16^3\times 32$ & 5000 & QCDSF\\
               5.2   & 0.1350 & $16^3\times 32$ & 8000 & UKQCD\\
               5.2   & 0.1355 & $16^3\times 32$ & 8000 & UKQCD\\
         \hline
               5.25  & 0.1346 & $16^3\times 32$ & 2000 & QCDSF\\
               5.25  & 0.1352 & $16^3\times 32$ & 8000 & UKQCD\\
               5.25  & 0.13575& $24^3\times 48$ & 1000 & QCDSF\\
         \hline
               5.26  & 0.1345 & $16^3\times 32$ & 4000 & UKQCD\\
         \hline
               5.29  & 0.1340 & $16^3\times 32$ & 4000 & UKQCD\\
               5.29  & 0.1350 & $16^3\times 32$ & 5000 & QCDSF\\
               5.29  & 0.1355 & $24^3\times 48$ & 2000 & QCDSF\\
         \hline
               5.4   & 0.1350 & $24^3\times 48$ & 1000 & QCDSF\\
      \end{tabular}
   \end{center}
   \caption{\footnotesize{Parameters for unquenched clover fermions 
            with Wilson glue.}}
   \vspace*{-0.30in}
   \label{table_uqC}
\end{table}
\begin{table}[ht]
   \begin{center}
      \begin{tabular}{l|l|l|l}
          $\beta$    & $am_q$ & Volume   & Confs. \\
         \hline
               8.45  & 0.140  & $16^3\times 32$ & $O(50)$ \\
               8.45  & 0.098  & $16^3\times 32$ & $O(50)$ \\
               8.45  & 0.056  & $16^3\times 32$ & $O(50)$ \\
               8.45  & 0.028  & $16^3\times 32$ & $O(50)$ \\
      \end{tabular}
   \end{center}
   \caption{\footnotesize{Parameters for quenched overlap fermions 
            with (tadpole improved) L\"uscher-Weisz glue.}}
   \label{table_qOLW}
   \vspace*{-0.30in}
\end{table}
(where for Wilson or clover fermions, $am_q = ( 1/\kappa - 1/\kappa_c)/2$
with $\kappa_c$ to be determined). Note that these refer to the number
of configurations used for finding the matrix elements;
for the nucleon/pion mass sometimes a higher statistic was used.


\subsection{Hadron masses}

\begin{figure*}[t]
   \hspace*{0.75in}
   \epsfxsize=11.00cm \epsfbox{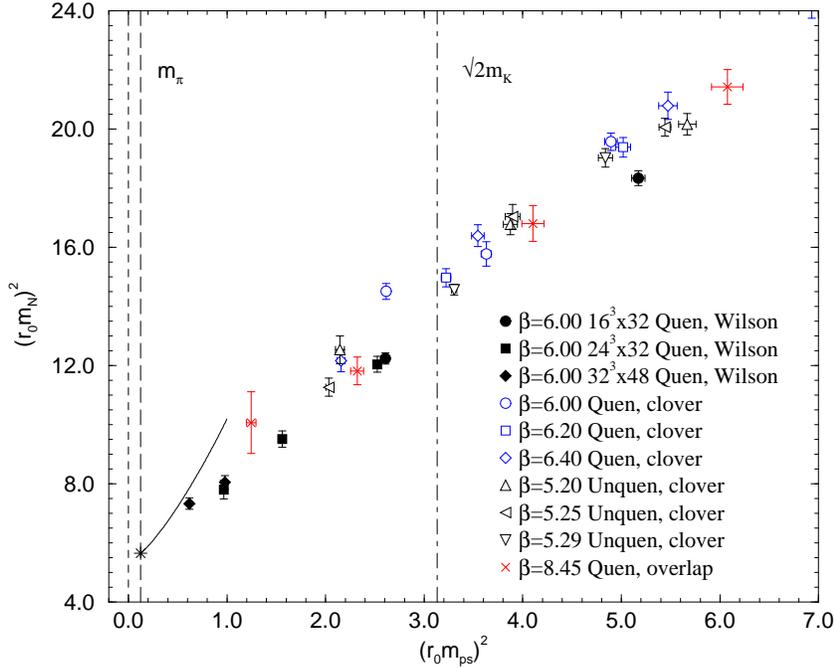}
   \vspace*{-0.30in}
   \caption{\footnotesize{$(r_0m_N)^2$ versus $(r_0m_{ps})^2$ for the
            four sets under consideration. The experimental nucleon mass
            is denoted by a star. The chiral limit $m_{ps}^2 = 0$
            is shown as a short dashed line, while the physical pion
            mass is denoted by the long dashed line. Also shown as
            a dot-dashed line is the mass of a hypothetical
            $\overline{s}s$ meson calculated as $\sim \sqrt{2}m_K$.
            The (quenched) $\chi$PT line uses the numerical values
            from eq.~(\ref{chiPT_numerical}).}}
   \label{fig_r0nuc2_r0pi2+gw}
\end{figure*}

To give some idea of where these parameter values lie, in
Fig.~\ref{fig_r0nuc2_r0pi2+gw} we plot the various nucleon masses
against the square of the pion mass. Also shown for comparison is the
position of a hypothetical $\overline{s}s$ pseudoscalar meson and
the physical pion, as well as the chiral limit.
The data lie in a narrow band tending to the physical nucleon
(denoted by a star in the plot). In general there do not appear
to be large discretisation effects. For the quenched data
(in particular) taking the $y$-axis to be $\propto m_N^2$
 seems phenomenologically to lead to linear lines (although possibly
the lightest point in a data set might be bending slightly upwards).
A pseudoscalar mass of $\sim 400\mbox{MeV}$ corresponds to
$(r_0m_{ps})^2 \sim 1.0$ while $m_{ps} \sim 600\mbox{MeV}$ gives
$(r_0m_{ps})^2 \sim 2.3$. Most of the lattice data lies above the region
$m_{ps} \sim 600\mbox{MeV}$, with the exception of the quenched Wilson
fermion results. In what region might the leading order $\chi$PT
be applicable? In Fig.~\ref{fig_r0nuc2_r0pi2+gw} we also plot
a representative curve (for the quenched nucleon mass).
A breakdown is seen at about $(r_0m_{ps})^2 \sim 1$ or
$m_{ps} \sim 400\mbox{MeV}$, as the gradient of the $\chi$PT result
is too steep. Higher order $\chi$PT might improve the situation
\cite{bernard03a,procura03a}.


\subsection{Nucleon matrix elements}

There are two distinct steps in determining nucleon matrix elements.
First the bare matrix element must be determined from the ratio
of three point nucleon-operator-nucleon correlation functions
to two point nucleon-nucleon correlation functions. Secondly this
matrix element must be renormalised. Techniques for both steps are
standard, eg \cite{gockeler95a}. In general, one calculates
non-singlet, NS, matrix elements because then the difficult to
compute one-quark-line-disconnected part of the matrix element cancels
(although for the vector current this additional piece is probably
very small, \cite{bakeyev03a}, so we shall ignore this point here).

For all the data sets for $v_2$ and the vector current the appropriate
renormalisation constant is known to one loop perturbation theory
(for overlap fermions the results are given in
\cite{galletly03a,horsley03a}). Either one can tadpole improve, TI, the 
result, for example TRB-PT, \cite{capitani01a,gockeler03a} or use
a non-perturbative method, NP, \cite{martinelli94a}. (For our variant
of the method see \cite{gockeler98a}.) For
overlap fermions and some of the unquenched clover results
this non-perturbative method has not yet been implemented,
so we must use TI-PT. As we wish to compare all our results, for consistency
we shall present them all here using TI-PT. The fact that the vector current
is conserved however allows a more direct NP determination of $Z_V$,
\cite{bakeyev03a,galletly03a}. For overlap fermions in particular this appears
to indicate that there are significant differences between perturbative
based results and NP results; however for Wilson or clover results
there is often only a couple of percent (or less) difference between
TI and NP results.


\section{MATRIX ELEMENT RESULTS}


\subsection{$v^{\NS}_2$}

\begin{figure}[t]
   \vspace*{0.05in}
   \epsfxsize=7.25cm \epsfbox{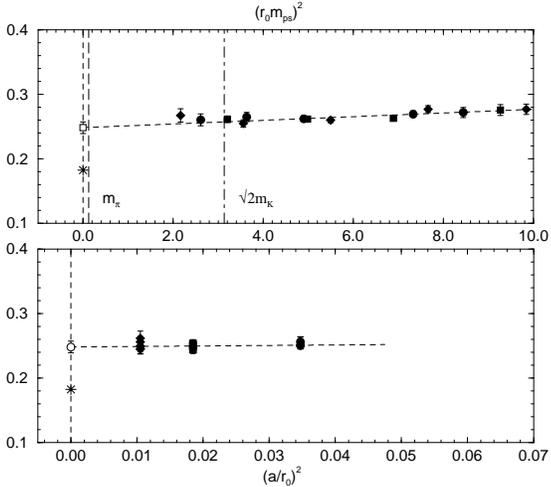}
   \vspace*{-0.30in}
   \caption{\footnotesize{
            $v^{\NS;\msbar}_{2b}- d_a (a/r_0)^2 - d_m ar_0m_{ps}^2$
            at a scale of $2\mbox{GeV}$ versus $(r_0m_{ps})^2$,
            upper picture, and 
            $v^{\NS;\msbar}_{2b} - (F_\chi^{v_2}(0)+ c(r_0m_{ps})^2)
            - d_m ar_0m_{ps}^2$
            against $(a/r_0)^2$, lower picture, for quenched clover
            fermions. Filled circles, squares and diamonds
            represent $\beta = 6.0$, $6.2$ and $6.4$ respectively.
            The MRS phenomenological value of $v^{NS;\msbar}_2$ is
            represented by a star.}}
   \vspace*{-0.30in}
   \label{fig_x1b_1u-1d.p0_020614_1307_lat03}
\end{figure}

We now show a selection of results. From the quenched clover fermion
data we shall first check for lattice spacing effects. In 
Fig.~\ref{fig_x1b_1u-1d.p0_020614_1307_lat03} we show the chiral and
continuum extrapolation for $v_{2b}$ (the `$b$' just denotes
the lattice represention used) using the formula given in
eq.~(\ref{practical}), together with a linear function in the
quark mass, eq.~(\ref{taylor}).
The upper graph represents the chiral physics and the lower graph
represents the discretisation errors. First we note that there seem
to be very small $a$ discretisation errors and indeed the numerical
effect of additional operators needed to ensure $O(a)$ improvement seems
to be negligible, \cite{gockeler03a}. Secondly a linear chiral
extrapolation seems to describe the data adequately. (Again, as for
the nucleon mass, we might expect $\chi$PT to be only valid for
$m_{ps} \lsim 400\mbox{MeV}$.) A result of
$v_{2b}^{\msbar}(2\mbox{GeV}) = 0.25(1)$ is found.
This is to be compared to the phenomenological result of $\sim 0.18$,
so we see about a $40\%$ discrepency.

\begin{figure}[ht]
   \vspace*{0.05in}
   \epsfxsize=7.25cm \epsfbox{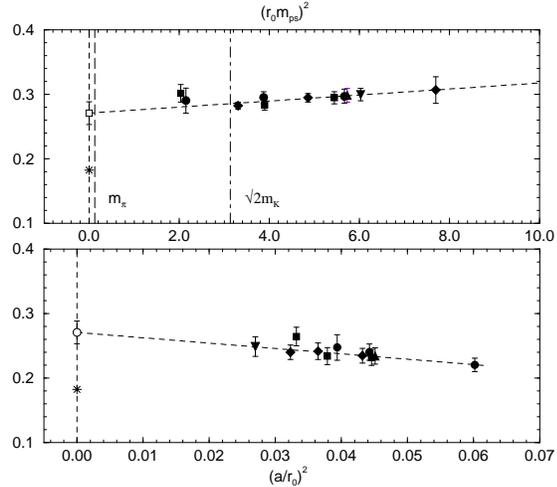}
   \vspace*{-0.30in}
   \caption{\footnotesize{Same as
            Fig.~\ref{fig_x1b_1u-1d.p0_020614_1307_lat03}
            for unquenched clover fermions. Filled circles,
            squares, up triangle, diamonds and down triangle
            represent $\beta = 5.20$, $5.25$, $5.26$, $5.29$ and
            $5.40$ respectively.}}
   \vspace*{-0.30in}
   \label{fig_x1b_1u-1d.p0_030710_1922}
\end{figure}

Turning now to the unquenched data, in
Fig.~\ref{fig_x1b_1u-1d.p0_030710_1922} we show the results.
Again similar conclusions hold as for the quenched case, although
it should be noted that we are not so close to the continuum limit.
We find $v_2^{\msbar}(2\mbox{GeV}) = 0.27(2)$. So at least in this
quark mass range quenching effects seem to be small.

\begin{figure}[ht]
   \vspace*{0.05in}
   \epsfxsize=7.25cm \epsfbox{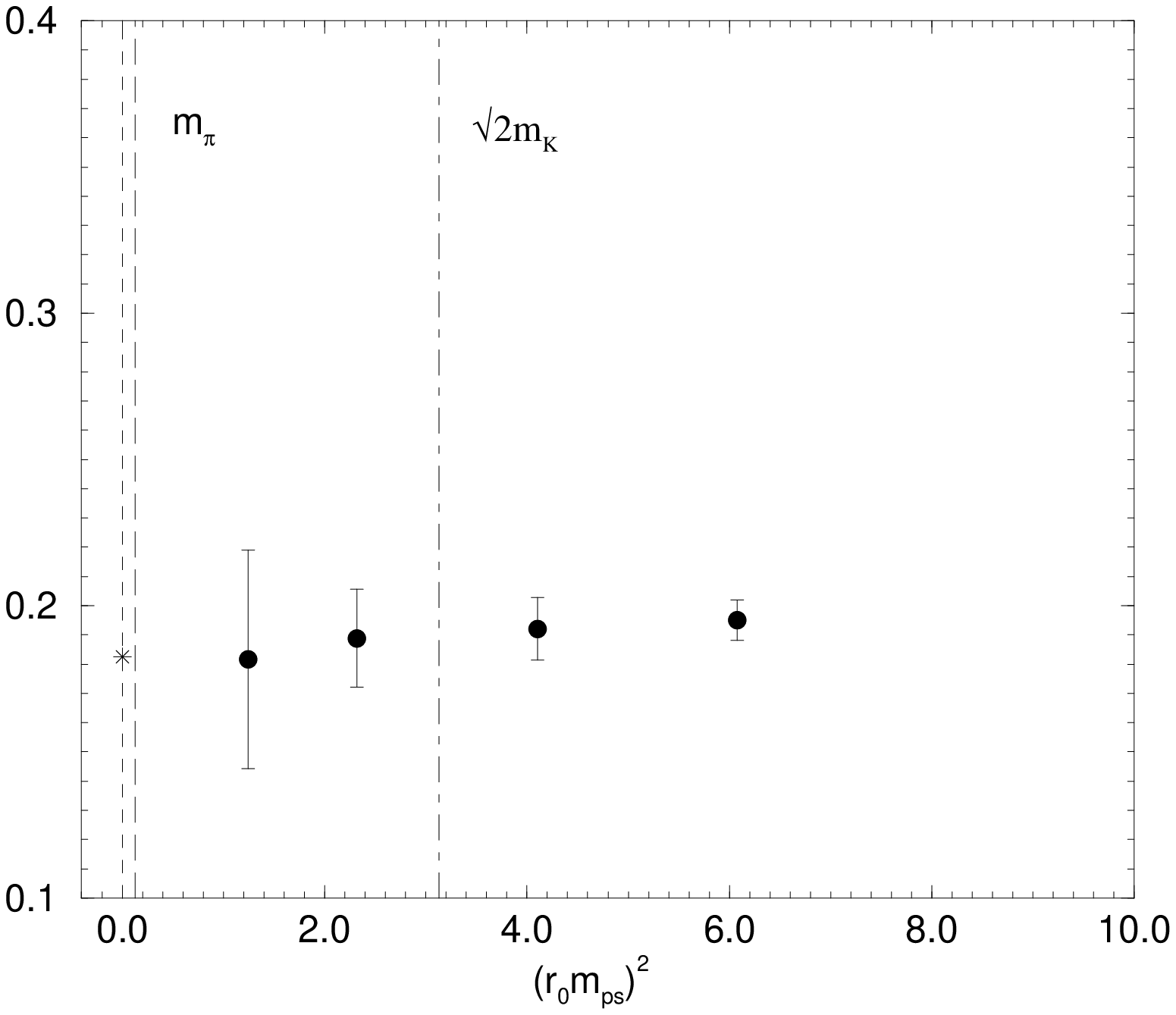}
   \vspace*{-0.30in}
   \caption{\footnotesize{$v^{\NS;\msbar}_{2b}(2\mbox{GeV})$ using
            quenched overlap fermions at $\beta = 8.45$.}}
   \vspace*{-0.20in}
   \label{fig_x1b_1u-1d.p0_040226_cairns03}
\end{figure}

Is the use of Wilson or clover fermions bad for the investigation of 
chiral properties and the chiral limit? Overlap fermions
having a chiral invariance on the lattice
in the massless limit, are in principle better.
In Fig.~\ref{fig_x1b_1u-1d.p0_040226_cairns03} we show
preliminary results for $v_2$ using quenched overlap fermions,
\cite{galletly03a}. Again we see the linear behaviour in the quark
mass confirmed. (The result is in surprisingly good agreement
with the phenomenological value. At present we have to use the
TI perturbation result for the renormalisation constant. As noted
previously when comparing a NP evaluation of $Z_V$ with a TI
perturbative evaluation the differences are greater than in the
Wilson or clover case. So perhaps a NP evaluation of the
renormalisation constant will give it a large value and so lead
to a larger result for $v_2$, more in line with other values presented
here.)

\begin{figure}[t]
   \vspace*{0.05in}
   \epsfxsize=7.25cm \epsfbox{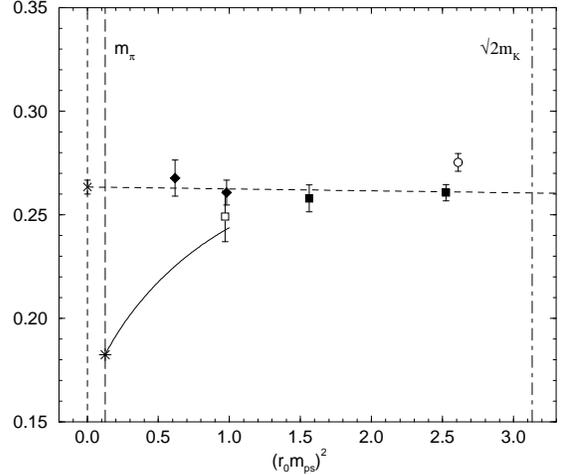}
   \vspace*{-0.30in}
   \caption{\footnotesize{$v^{\NS;\msbar}_{2b}(2\mbox{GeV})$ using
            quenched Wilson fermions at $\beta = 6.0$. The $16^3\times 32$
            results are shown as circles, the $24^3\times 32$ results
            as squares and the $32^3\times 48$ results as diamonds.
            A linear fit, eq.~(\ref{taylor}), is also shown as a dashed line.
            The (quenched) $\chi$PT line uses the numerical values
            from eq.~(\ref{numerical_c2}).}}
   \vspace*{-0.20in}
   \label{fig_x1b_1u-1d.p0_030624_1737+chipt}
\end{figure}

Finally we investigate the chiral limit, using light Wilson quarks
on large lattices. In Fig.~\ref{fig_x1b_1u-1d.p0_030624_1737+chipt}
we show the results for $(r_0m_{ps})^2$ between about $0.5$ and $2.5$ or
$m_{ps} \sim 300$ -- $600\mbox{MeV}$. First it seems that the numerical
values are about the same as for the clover quenched case,
which might indicate that finite lattice effects are again small.
All the results are again rather flat (and this continues to heavy
quark masses \cite{gockeler95a,capitani01a}). Also shown
is a linear fit from eq.~(\ref{taylor}) together with $\chi$PT results,
eq.~(\ref{chiPT}), starting from the MRS phenomenological value and
using the quenched result in eq.~(\ref{numerical_c2})
for chiral scale $1\mbox{GeV}$. (Using larger values of $c_1^{v_2}$
leads to a steeper slope, while using a smaller value for $\Lambda_\chi$
reduces the slope.)
For the present numerical results to reach the phenomenological
value seems difficult, as down to $\sim 300\mbox{MeV}$ they seem to
be rather linear. Of course we would not necessarily expect complete
agreement with the phenomenological result, however past experience
with the quenched approximation does suggest that it can lead to
values only a few percent away from experimental values.


\subsection{Form Factors}

\begin{figure}[t]
   \vspace*{0.05in}
   \epsfxsize=7.25cm \epsfbox{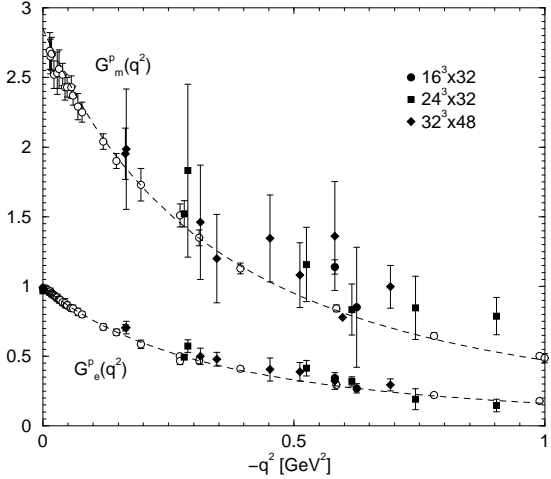}
   \vspace*{-0.30in}
   \caption{\footnotesize{$G^p_e$ and $G^p_m$ for the proton
            as functions of $-q^2$ using quenched Wilson fermions
            at $\beta = 6.0$. The previous experimental results are given
            by the open circles.}}
   \vspace*{-0.20in}
   \label{fig_Ge+Gm_Vff_Qu-Qd_GeV2_b6p00_wil}
\end{figure}

\begin{figure}[t]
   \vspace*{0.10in}
   \epsfxsize=7.75cm \epsfbox{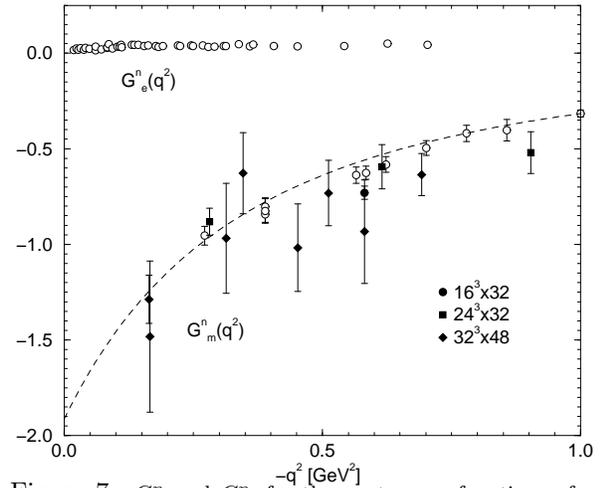}
   \vspace*{-0.50in}
   \caption{\footnotesize{$G^n_e$ and $G^n_m$ for the neutron
            as functions of $-q^2$ using quenched Wilson fermions
            at $\beta = 6.0$.}}
   \vspace*{-0.20in}
   \label{fig_Ge+Gm_Vff_Qd-Qu_GeV2_b6p00_wil}
\end{figure}

Finally we present our results for electromagnetic form factors, as
given in eq.~(\ref{formfactor}). We have $F_1(0) = 1$ (charge conservation
leading to a NP evaluation of $Z_V$ on the lattice) and $F_2(0) = \mu - 1$,
the anomalous magnetic moment in magnetons. It is usual to give the
results for the Sachs form factors
\begin{eqnarray}
   G_e(q^2) &=&
      F_1(q^2) + {q^2\over (2m_N)^2} F_2(q^2) \,,
                                                \nonumber \\
   G_m(q^2) &=& F_1(q^2) + F_2(q^2) \,.
\end{eqnarray}
Previous experimental results give phenomenological dipole fits,
for the proton, $p$ or neutron, $n$,
\begin{eqnarray}
   G_e^p(q^2) &\sim& {G_m^p(q^2) \over \mu^p}
                              \quad \sim \quad
                              {G_m^n(q^2) \over \mu^n}
                                                \nonumber \\  
                          &\sim&
                            { 1 \over
                            \left(1+ {-q^2 \over m_V^2}
                            \right)^2 } \,,
\end{eqnarray}
and
\begin{eqnarray}
   G_e^n(q^2) &\sim& 0 \,,
\end{eqnarray}
with $m_V \sim 0.82 \mbox{GeV}$, $\mu^p \sim 2.79$ and
$\mu^n \sim -1.91$. (Present experimental results indicate however
that, \cite{jones99a}, $G^p_e(q^2) /G^p_m(q^2)$ is decreasing
with increasing $-q^2$.) In Fig.~\ref{fig_Ge+Gm_Vff_Qu-Qd_GeV2_b6p00_wil}
we show $G^p_e$ and $G^p_m$ for the proton and in 
Fig.~\ref{fig_Ge+Gm_Vff_Qd-Qu_GeV2_b6p00_wil} for the neutron. Note that we
have made simple linear chiral extrapolations here, see eq.~(\ref{taylor}),
but at present we can do little more than this, as in particular on the
large $32^3\times 48$ lattice, we only have two quark mass values. The
technique used here is the same as described in
\cite{capitani98a,gockeler03c}. Previous lattice calculations described 
in these papers seem to show little or no lattice artifacts.
Due to the larger lattice size, we are now able to go to smaller momentum
transfer $-q_{min}^2 \sim (2\pi/32a)^2 \sim 0.17\mbox{GeV}^2$
than was previously possible. We find that the lattice result tracks
the experimental values reasonably well, although with rather large errors.
Finally we note that as usual the electric form factor of the neutron
remains difficult to measure on the lattice (see \cite{tang03a}
for a recent attempt).


\section*{ACKNOWLEDGEMENTS}

The Wilson and clover numerical calculations were performed on the
Hitachi {\it SR8000} at LRZ (Munich), the Cray {\it T3E}s at
EPCC (Edinburgh), NIC (J\"ulich) and ZIB (Berlin)
as well as the APE/Quadrics at NIC (Zeuthen).
The overlap numerical calculations were performed at NIC J\"ulich,
HLRN Berlin and NeSC Edinburgh (IBM Regattas),
NIC Zeuthen and Southampton (PC Clusters), and HPCF Cranfield (SunFire).
We thank all these institutions for support. This work is supported
by the European Community's Human Potential Program under contract
HPRN-CT-2000-00145 Hadrons/Lattice QCD and by the DFG 
under contract FOR 465 (Forschergruppe Gitter-Hadronen-Ph\"anomenologie).




\begin{thebibliography}{99}

\bibitem{gockeler02a}
   M. G\"ockeler et al.,
   Lattice 2002,
   Nucl. Phys. Proc. Suppl. 119 (2003) 398, hep-lat/0209111.

\bibitem{labrenz93a}
   J. Labrenz et al.,
   Lattice 93,
   Nucl. Phys. Proc. Suppl. 34 (1994) 335, hep-lat/9312067.

\bibitem{bernard93a}
   V. Bernard et al.,
   Z. Phys. C60 (1993) 111, hep-ph/9303311  .

\bibitem{detmold01a}
   W. Detmold et al.,
   Phys. Rev. Lett. 87 (2001) 172001, hep-lat/0103006.

\bibitem{arndt01a}
   D. Arndt et al.,
   Nucl. Phys. A697 (2002) 429, nucl-th/0105045.

\bibitem{chen01a}
   J. Chen et al.,
   Phys. Lett. B523 (2001) 107, hep-ph/0105197.

\bibitem{chen01b}
   J. Chen et al.,
   Nucl. Phys. A707 (2002) 452, nucl-th/0108042.

\bibitem{bernard03a}
   V. Bernard et al., hep-ph/0307115.

\bibitem{procura03a}
   M. Procura et al., hep-lat/0309020.

\bibitem{gockeler95a}
   M. G\"ockeler et al.,
   Phys. Rev. D53 (1996) 2317, hep-lat/9508004.

\bibitem{bakeyev03a}
   T. Bakeyev et al.,
   hep-lat/0305014.

\bibitem{galletly03a}
   D. Galletly et al.,
   Lattice 2003, hep-lat/0310028.

\bibitem{horsley03a}
   R. Horsley et al.,
   in preparation.

\bibitem{capitani01a}
   S. Capitani et al.,
   Lattice 2001,
   Nucl. Phys. Proc. Suppl. 106 (2002) 299, hep-lat/0111012.

\bibitem{gockeler03a}
   M. G\"ockeler et al.,
   in preparation.

\bibitem{martinelli94a}
   G. Martinelli et al.,
   Nucl. Phys. B445 (1995) 81, hep-lat/9411010.

\bibitem{gockeler98a}
   M. G\"ockeler et al.,
   Nucl. Phys. B544 (1999) 699, hep-lat/9807044.

\bibitem{jones99a}
   M.~K. Jones et al,
   Phys. Rev. Lett. 84 (2000) 1398, nucl-ex/9910005.

\bibitem{capitani98a}
   S. Capitani et al.,
   Lattice 98,
   Nucl. Phys. Proc. Suppl. 73 (1999) 294, hep-lat/9809172.

\bibitem{gockeler03c}
   M. G\"ockeler et al.,
   hep-lat/0303019.

\bibitem{tang03a}
   A. Tang et al., hep-lat/0307006.

\end{thebibliography}
\end{document}